# DISCOVERING THE IMPACT OF KNOWLEDGE IN RECOMMENDER SYSTEMS: A COMPARATIVE STUDY

Bahram Amini[1], Roliana Ibrahim[2], Mohd Shahizan Othman[3]

[1,2,3] Faculty of Computer Science and Information Systems,
Universiti Teknologi Malaysia (UTM), 81310 Malaysia
`avbahram2@live.utm.my, roliana@utm.my, shahizan@utm.my`

## ABSTRACT

*Recommender systems engage user profiles and appropriate filtering techniques to assist users in finding more relevant information over the large volume of information. User profiles play an important role in the success of recommendation process since they model and represent the actual user needs. However, a comprehensive literature review of recommender systems has demonstrated no concrete study on the role and impact of knowledge in user profiling and filtering approache. In this paper, we review the most prominent recommender systems in the literature and examine the impression of knowledge extracted from different sources. We then come up with this finding that semantic information from the user context has substantial impact on the performance of knowledge based recommender systems. Finally, some new clues for improvement the knowledge-based profiles have been proposed.*

## 1. INTRODUCTION

The rapidly growing amount of information on the Web provides so much difficulty for internet users to find desired information. This case may be even worse if users don't utilize appropriate search tools for information discovery. In recent years, Web personalization has received much attention to help Internet users with the problem of information overload [1], i.e. users often receive huge amount of unrelated information resulting from the Web search. As the volume of content and the number of services on the Web rapidly grows, Web personalization and recommender systems (RS) are more than before in demand.

Recommender systems offers new unvisited items such as goods, web pages, etc based on the user's preferences maintained in the user's profile. Recommender systems deal with the information overload problem by filtering items or services that potentially may match the user's interest. They help users efficiently overcome the problem of content overload by filtering irrelevant information when users search for desired information. In traditional recommender systems, the user profile is keyword-based which, to some extent, involves only those items that match specific keywords in the user's preferences [2]. However, many problems with keyword-based profile representation methods arises such as losing a lot of useful information, inaccuracy in the recommendation items, and user dissatisfaction. One approach to overcome these problems is to extracting and utilization semantic-based information from the domain and incorporating them within the stages of the personalization process.

The Semantic Web mechanisms, however, incorporate additional knowledge about the user's preferences into recommendation process, specifically in the user profile, by elicitation information from the pages user visited and rated during his practice with the system, and therefore provides more accuracy and flexibility to the personalization processes [3]. Moreover, since Semantic Web teknologies enables to attain more accurate insight into the meaning of the underlying information, it helps to develop semantically enrich descriptions of user interests for further improvement of Web personalization techniques [4]. Furthermore, with the over growing advances in the semantic web field, the possibilities for improvement of recommendation process by engaging more intelligent approaches encourage scholars [5].

In this paper, we investigate the latest efforts for recommender systems on the basis of engaging knowledge sources, i.e. the origin of the knowledge employed to generate recommendations and the impact of the knowledge on the recommendation efficiency. In some approaches this knowledge comes from the knowledge of other users' preferences while in the other approaches it comes from ontological or inferential knowledge about the domain, provided by knowledge engineering tasks. We also compare some representative approaches from the perspective they employ knowledge in the recommendation process and come up with potential improvements in the context.

The rest of the paper is organized as follows: A quick review of related works is presented in Section 2. We describe the fundamental approaches of recommender systems in Section 3. The analysis and evaluation of several approaches as well as some semantic based recommender systems are discussed in Section 4 and Section 5 respectively. The challenges and potential improvements to user profiles are described in Section 6. We discuss the result and our contribution in Section 7.

## 2. RELATED WORK

Uchyigit [2] surveys the state-of-the-art of the techniques which are used to semantically enhance user modeling within the recommendation phases. This work introduces the web usage mining as an approach to capture and model the user's behavioral patterns as user interacts with the system. Dai et al., [6] overview the approaches for incorporating semantic knowledge into Web usage mining and personalization processes including Content Features and Domain Ontologies. Content Features include keywords, phrases, category names, or other textual content embedded as meta-information in content of web pages. In contrast to content features, domain ontology represents the knowledge contained in the web site which makes it possible to have a global architecture of common objects, their properties, and their relationships in the domain of discourse.

Sieg et al., [7] review an approach to personalized web search which incorporates models of "user context" based on ontological profiles by assigning implicitly derived user's interest scores to concepts in the prebuilt domain ontology. They define "context" as the immediate and past user activities as well as knowledge from a pre-existing ontology as an explicit representation of the domain of interest. Cantador et al., [5] investigates the unification of personal information with ontologies using the contemporary knowledge representation methods associated with Web2.0 applications, such as community tagging. They propose a method for the unification of tags, grounding to a shared representation in the form of WordNet and Wikipedia, with ontologies. They incorporate individuals' tagging history into the user ontological profiles by associating tags with the ontology concepts.

Mobasher [8] distinguishes the Web personalization approaches in three general categories based on their approaches: knowledge-based, content-based, and collaborative-based systems. These approached are classified from architectural, algorithmic and the way Semantic Web teknologies have been exploited in augmenting information to the user profile. Here, knowledge refers to the information about items, users (such as demographic characteristics) through ontologies, constraints (such as rule-based methods), folksonomies (social tags that add meta data to shared contents)[9], etc.

## 3. EFFECTS OF KNOWLEDGE IN RECOMMENDER SYSTEMS

In the literature, personalization systems are classified in different ways. Some researchers have categorized them based on decision modelling or utility functions [10]. We include the discussion of demographic recommender as the basic approach which augments domain knowledge into the RS process. We also describe hybrid approach [11] as a possible combination of multiple approaches including content-based, collaborative-based, and knowledge-based approaches. In the following, the basic RS techniques and the effects of applying contextual knowledge in recommender systems are discussed.

### 3.1. Knowledge based recommender

Knowledge based or so called rule-based personalization systems recommend items by generating manually or automatically a number of decision rules. Knowledge-based recommenders emphasizes on explicit domain knowledge about the items as well as implicit knowledge about the users (such as psychographic, demographic, or other personal attributes of users) to extract pertinent recommendations [12]. Such systems rely on development of knowledge intensive rules that are used to propose items to users which exactly match with the specification identified by the rules. Manually development of rules, like most rule-based recommender systems, relies heavily on knowledge engineering techniques for development the rule base correspond to the specific attributes of the target domain. Therefore, the user profiles are developed through direct interactions of the users with the system. Some research, however, have focused on automatically deriving decision rules by employing machine learning approaches for classifying users into several classes based on their demographic characteristics, and therefore, the classification can be used for personalization [13]. For example, many e-commerce applications that provide item recommendation employ manually generating rule-based approaches to offer more interesting products or services to their customers. In this case, web site administrator creates the decision rules manually based on, for example, personal interests of users.

However, manually generated decision rules for such personalization systems are associated with different drawbacks. Firstly, the system suffers from bottleneck problem in the knowledge engineering process. Essentially, the important part of personalization process is depended on the task of knowledge engineering. It is performed by the system developer who manages the rule base by selecting specific attributes of the domain. Secondly, the methods of user profile development suffer from inherit problems; during the interactions of users with the site, user profiles are generally created. For classifying users and extracting decision rules, developer mainly focuses on machine learning as a means of rule generation. Therefore, the input is subjective and affected by the description of user's interests given by the users. Moreover, the user profile is not updating and therefore performance of the personalization process declines over the time.

### 3.2. Content-based recommender

Traditional content-based recommendation approaches generally drive the user's profile from the contents of Web pages that users visited or rated. It then compares the information from the new Web pages with the user's profile and, those items from the Web pages which are most similar to user's profile are recommended. Actually, it assumes that the new choices of a user are very similar to the choices made by him in the recent past. Generally, the description of items in the user profile includes a set of attributes that identifies the items user is interested in. In other words, description of the items user was previously seen and rated builds the user profile. In recommendation process, characteristics of unseen pages with the features stored in the user profile are compared and the items that are similar enough to the features stored in the user profile are recommended [13].

In many Web-based personalized applications such as e-commerce and e-learning sites, several techniques for document modeling, information filtering, and techniques for deriving information from the pages content are proposed. In such application, user profiles are generally described as vectors so that every entry of vectors represents a weight or an interest degree of each item in the Web pages. For a particular item, predictions about the user interest can be computed by comparison of vectors by employing Cosine similarity [4] or Bayesian classification [13] approaches.

However, content-based filtering approaches suffer from different drawbacks. The first deals with the way of creating user's profile. In fact, user profiles are extracted from text that describes items which previously rated by the users. In this method, user profile provides overspecialized problem by missing important implicit relationship between the objects within the Web pages [14]. For instance, common features between objects in a particular context such as item utilization may be lost. Besides, the system is strongly dependent on the availability of content for describing items being recommended [15]. Moreover, this method may offer lots of relevant and irrelevant information, mainly because each item which matches with some keywords will be considered interesting to the user regardless of its context. For example, if the word player exists in a page about football games (from the sport context) then a web page about agent player (from the artificial intelligence context) will be also recommended. In order to overcome such shortcomings, it is important to model the semantic meaning of the data in the domain of problem. In recent years, ontologies and the Semantic Web teknologies have been extremely used to achieve this functionality. Ontologies, as an explicit description of domain concepts, represent more precise and less ambiguous domain of concern than usual keyword-based approaches [2].

### 3.3. Collaborative-based recommender

Collaborative-based or so called social-based are an alternative approach to the previous approaches, aiming to improve the limitations of content-based approach. It exploits the other user's profiles in the same community and recommends new items not previously rated or seen by the user based on the assumption that similar users have similar interests in the same community. Therefore, recommendations take places based on the user similarity and recommend items from the interesting list of other people in the same community. Recommendation is therefore achieved by finding common characteristics in the preferences of other users in the community which are expressed in the form of item ratings, maintained in their user profiles. To perform recommendation, each user is associated with a group of nearest neighbour users by comparing the user's profile with the other user's profiles. Then, items in the candidate user's profile act as a template for the target user, and can be recommended to the target user. However, when the percentage of users compared to the content is significant small, only a few numbers of the items will be accessed by the target user. Accordingly, the spare coverage of ratings [16] produces a sparse set of recommendable items.

The k-Nearest-Neighbor (k-NN) approach which is a standard memory-based classification method has been engaged as one of the fundamental techniques to fulfil the recommendation task. This approach simply compares the active user profile with the other user's profiles in the community by computing the top most k users who have similar preferences to the current user. However, the k-NN approach imposes a considerable limitation to the collaborative filtering and suffers the lack of scalability [13]. Because, when the number of items and users increases, it often leads to less acceptable latency state for preparing recommendations. Another weakness is the sparsity of the available Web data; as the number of items increases, the percentage of user rating decreases. It is mainly because only a small number of rating values relevant to the rated items are incorporated.

There are also two additional disadvantages associated with the collaborative filtering approach. Firstly, the ratings of all items should be provided before commencing the recommendation task. This is referred as the new item rating problem [15]. Secondly, new users have to perform rating of certain number of items before obtaining appropriate recommendations from the system, which is referred as the new user problem [1].

In order to mitigate aforementioned problems, a number of improvement approaches have been developed; dimensionality reduction approach [17] in order to mitigate the sparsity problem of data as well as the offline categorization technique [1] of user records through clustering methods. Another approach to improve collaborative filtering is to integrate additional information from other sources such as user's demographics data [18]. A variant of the traditional collaborative approach which is called model-based collaborative filtering has been proposed to enhance the traditional collaborative filtering approach. The item-based collaborative filtering, a variant of model-based collaborative filtering, starts from the same user rating profile databases and builds an item-item similarity matrix in offline situation. Each entry in the matrix expresses the similarity among pairs of the considered items. Those similarity items are then used in the prediction phase to generate recommendations. The item-item similarity is based on the user ratings. Each item is represented in the form of an n-dimensional matrix where 'n' refers the number of users. Techniques such as Cosine similarity or the Correlation based similarity [19] has been used to discover the similarities between pairs of items. In online prediction phase, the item similarity matrix is used to generate recommendation of not rated items. The values in the matrix are the sum of weighted rating of neighboring items which have previously been rated by the current user. As the number of items in the similarity matrix increases, it naturally requires big amount of memory space.

## 3.4. Demographic recommender

A demographic recommender system provides recommendations based on a user's demographic profile which involves user's demographic data such as gender, age, date of birth, education, and other personal features [11]. This approach categorizes users into groups based on their demographic characteristics and recommends objects accordingly [20]. More precisely, it assumes users in the same category have the same taste or preferences. Recommendations are issued for new users by first identifying the category user belong to and then by locating preferences of other users in the same category.

However, demographic recommender is dependent to availability of demographic data to work well. Because of many limitation and difficulty to extract demographic knowledge, this approach is not applicable in most domains including personal search and education. According to this limitation, few systems use demographic data within the recommendation process [15]. Even that data are being collected on the web, they tend to be of poor quality. Moreover, recommendations based on demographic data have been shown to be less accurate than those based on the content and user collaboration [13].

## 3.5. Hybrid approach

Hybrid approach combines two or more techniques described earlier in different ways to improve recommendation performance in order to tackle the shortcoming of underlying approaches including cold-start or data sparsity problem. For example, a knowledge-based and a collaborative system might be combined together to achieve more robust recommender system than the individuals components[8]; knowledge-based component can overcomes the cold-start problem by generating recommendations to new users whose profiles are small, and the collaborative module can help by finding those users who have similar preferences in the domain space that no knowledge engineer could have predicted.

However, current hybrid approaches still suffer from a few drawbacks [21]. First, there are insufficient contextual information to model users and items and therefore weaknesses to predict user taste in domains with complex objects such as education. Second, there is no support for multi criteria rating which requires user make judgment based on several factors such as quality and user orientation. For example, in some applications such as e-learning, users require to rate an objects regarding the study time and individual background knowledge simultaneously. Lastly, there is also the limitation in nearest neighbour based computing, scalability problem, since the computation time grows rapidly with the number of users and objects [22].

The most common approaches that have widely been used in hybrid approach are "content-based" and "collaborative-filtering". Moreover, hybrid recommenders can also be classified based on their operations into seven different types including weighted, switching (or conditional), mixed, feature-based (property-based), feature combination, cascade, and meta-level. Interested readers can refer to [11] and [19] for further discussion of hybridization approaches.

## 4. ANALYSIS OF RECOMMENDER SYSTEMS

In the context of collaborate-based and content-based approaches, a number of methods have been explored the impact of integrating knowledge to the user profile and demonstrated the potential improvement to the recommendation process [8]. Such approaches utilize the knowledge acquired from the context through the modelling of the domain or user's behaviour in the form of ontological knowledge and reasoning utilities. In this scenario, we evaluate the recommender systems on the basis of augmenting extensive knowledge in the recommendation process and synthesis their impacts on the recommendation process as well as drawbacks affected by the lack of sufficient knowledge.

Table 1 summarizes and compares the basic recommendation approaches based on the knowledge impression. As shown, insufficient knowledge in collaborative filtering produces as many shortcomings as knowledge-based approaches because they rely on the knowledge obtained solely from the user interaction with the system but not their contextual or implicit knowledge from the domain. Similarly, the external knowledge provided by domain expert in knowledge-based approach does not produce critical problems and therefore enhance the recommendation performance and system usability. In other words, performance can be improved by integrating enriched knowledge from the context or exploiting technical approaches for updating user profile. This is the reason why knowledge-centric approaches such as ontology-based, web usage mining and formal concept analysis are widely used in recommendation systems for enriching the recommendation process [1]. The main tasks of such techniques are discovering and extracting as well as integrating the domain knowledge accompanying with user knowledge into the recommendation process [20].

On the other hand, augmenting knowledge into the recommendation process, as shown by demographic-based approaches, does not necessarily mean good performance and high accuracy. In fact, appropriate knowledge from the context, i.e. the content, user behaviour, user modelling should be incorporated into recommendation process in order to improve the total performance and tackle the drawbacks. Therefore, factors such as the type, structure, and adaptability of knowledge to the framework, either contextual or non contextual, have direct impact on the recommendation process.

Moreover, we explored three general approaches for incorporating knowledge into the recommender systems including modelling, similarity computing (filtering), and re-ranking. Table 2 summarize those approaches that have been used for extraction, presentation, and reasoning the intensive knowledge in respected recommender systems which all deal with integrating knowledge with recommendation process. Modality which is either implicit, explicit, or both indicates how the corresponding component manipulates the relevant knowledge in the RS process. As an example, in Web usage mining [24], the process of knowledge extraction is implicit because it deals with user behaviour stored in log files while in cosine similarity computing that is used for filtering processes, the knowledge maintained in the underlying ontology is explicit. Re-ranking deal with the task of arrangement the most relevant results on the topmost by bringing the result as relevant as possible to the user preference in descending order [21] which, in turn, increases user satisfaction.

Table 1: The comparison of recommender approaches based on the knowledge impression

| Recommend Approach | Source of Knowledge | Type of knowledge | K. extraction method | Drawbacks |
|---|---|---|---|---|
| **Knowledge based** | psychographic, demographic, personal attributes of users | Decision Rules | Machine-learning, K. engineer interaction | Bottleneck in K. engineering, subjective user profile, static user profile |
| **Content based** | contents of web pages | description of items in the user profile (a set of attributes identifying the items), item-item relationship | document modeling, information filtering, information extraction | overspecialized problem, dependent on the availability of content, syntax-based recommendation (losing semantic meanings) |
| **Collaborative based** | other user's profiles (interesting list of other users in the community) | similarity matrix (shared features of other users' preferences in the community) | K-Nearest Neighbor (kNN), Cosine or Correlation based similarity | spare coverage problem, latency state problem, sparsity problem, new item rating problem, new user problem, cold-start problem, violate user privacy |
| **Demographic based** | Users' demographic data such as gender, age, date of birth, education, etc | Category membership, | Classification methods, locating group interests | dependent to availability of demographic data, less accurate (poor quality of demographic data) |

## 5. OVERVIEW OF THE RECOMMENDER APPROACHES

In this section, we present one sample of the state-of-the-art of recommendation systems in each category and discuss the basic approaches reviewd previously. We then highlight their exclusive methodologies for resolving the drawbacks of underlying approach with the aid of knowledge, extracted from the respected domain.

Table 2: Three kind of augmenting knowledge into recommender systems

| RS component | Presentation Approach | Extraction methods | Modality |
|---|---|---|---|
| Context Modelling | Semantic Web Teknologies, Decision tree, tagging | Text analysis [22], Machine learning methods [23] (Classification, Clustering, Association analysis, decision tree, Neural Networks) | Implicit /Explicit |
| | Association Rule, Sequence Pattern, Semantic information (ontology), weighted concepts | Web usage mining [24] (clustering, association rule mining, and sequential pattern discovery), fuzzy methods [25] | Implicit |
| Matchmaking and Filtering | Keyword-based, Ontology-based, Graph-based, Similarity matrix | Cosine similarity, k-NN, Pearson and Spearman Correlation, Mean-squared [15], Bayesian classification, Correlation based similarity | Explicit |
| Re-ranking | Descriptive feature, Vector Space Model [26] | Rule-based, item-rating (weighting, scoring), time-related, probabilistic inference [27], Google's PageRank [28] | Implicit |

The work presented in [14] is a content-based approach that relies on the main strengths of the content-based personalization paradigm, aiming to overcome the overspecialized problem. It addresses the challenges of TV viewers which faces information overload problem and the overwhelming of interaction yielded by the digital receivers. To assist viewers, it employs a reasoning-based approach which filters those programs that only match the viewers' preferences from thousands of TV programs. This approach offers diverse programs without relying to other users' preferences and preserving other user's privacy. Specifically, this approach resolves the syntactic restrictions of the typical content-based method by putting two reasoning techniques on which derived from the Semantic Web technology: Semantic Association (SA) and Spreading Activation techniques [14]. These techniques abandon the traditional syntactic similarities, but look for relationships between the user's preferences and the formalized items previously annotated by the domain ontology. Both of these techniques efficiently discover semantic relationships between items and explore new knowledge about the users' interests. Applying this knowledge enables the recommendation system offers more accurate items in a more flexible manner.

In [29] a semantic framework using the textual analysis for personalized advertisement has been proposed. The framework exploits ontology-based user profiles in order to tackle the vocabulary impedance [30] as well as the cold-start problem in the recommendation task [31]. In order to tackle the first problem, semantic data are retrieved from the content in the domain and expressed in a vocabulary or domain ontologies. A combination of linguistic analysis and exploiting of lexical graphs, the task of classifying text into ontology concepts has been accomplished. The second problem has been tackled by the reference ontologies derived from the basic information of the domain. This approach improves accuracy and completeness of recommendations by employing explicit domain knowledge represented by ontology or taxonomy. Therefore, for each ontology concept, a concept vector containing bag-of-words is constructed by creating an index training set of web pages. Alternatively, advanced inference methods such as reasoning over the domain ontologies and technique to implement fuzziness and weighting user preference are employed.

This approach also employs semantic knowledge and statistical analysis for terminology extraction from the body of advertisements a user visited or rated. Eventually, it filters and re-ranks ads according to the user's score stored in the user's profile. When the user consumes the web page contents (ads, articles, etc), the textual data of the items are linguistically analyzed and its semantic information is extracted. This information is translated into a set of user preferences or semantic user profile. Then, user interests are semantically compared and matched to a set of candidate Web pages that are annotated to recommend ads with high confidence degree.

OntoCopi [32] is a recommendation system which determines communities of practices (COP) by analyzing ontologies of the relevant domain and considering common features of the community such as who attended the same event, who co-authored the same paper and who worked on the same project, and so on. An example of a COP might be a number of working people in an organization who perform the same or similar jobs and have many practices in common. All of these informal relation are being discovered and being considered in the user profile. OntoCopi uses an existing ontology, AKT, for identifying group of users in the research community. It recommends researchers appropriate papers based on the similarity of the researcher's profile with the available papers. An interesting feature of the system is that it determines communities of common interests by processing relations among concepts of ontology including event attending, authorship, supervision, and research interest, etc.

The approach presented in [33] is a hybrid recommender system for e-learning environment which integrates the recommendation utility into a scientific repository, HyperManyMedia. HyperManyMedia is an online learning system which contains educational content of courses, lectures, multimedia resources, etc and enables learners to search the repository in terms of keywords through a search engine interface. This recommender is driven by two basic recommendation components: content-based component which engages the domain ontology model and the rule-based component which is a learner's interest- based and cluster-based module. The domain ontology model encompasses the learning materials including concepts and sub-concepts of colleges, courses, and lectures as a hierarchy. Similarly, the learner's ontology represents a subset of the domain ontology e.g., an ontology containing a personalized (or a pruned subset of) the whole domain which is a hierarchy of college, courses, lectures that the learner is interested in. Moreover, the combination of such knowledge enriched content-based and rule-based approaches which exploit different weights of concepts/sub-concepts influences the ranking of the documents retrieved from the online repository.

Rule-based component uses clustering techniques to extract functional/descriptive features from the learning corpus, features that have been added to the corresponding concepts in the hierarchy of domain ontology. There is also a semantic mapping between the learners' query and the corresponding semantic profiles which presents their learner interests. Each underlying component has been influenced by the re-ranking strategy, applied with different factors on the retrieved documents from the corpus. The experiments with the two common measurements of recommender systems, e.g., top-n recall and top-n precision, measured the effectiveness of re-ranking process based on the learner's semantic user profile and proved the effectiveness of the proposed approach.

Table 3 summarizes the key information of these approaches and contrasts the effectiveness of contextual knowledge as a panacea of aforementioned problems. However, some problems still exist and therefore demand new kind of knowledge to promote the recommendation and overcome the drawbacks. For example, approaches that employ domain ontology is prone to cold start problem unless they employ an updating strategy for incorporating over changing user interests.

Table 3: The comparison of recommender approaches based on the knowledge impression

| Approach | Domain | Knowledge | Analysis | Drawback |
|----------|--------|-----------|----------|----------|
| Y. Blanco-fernández, et al: Content based | TV programs | Reference ontology + formalized item annotation | Reasoning+ Semantic Association and Spreading Activation techniques | Neutral to user's moment and mood, static user profile |
| D. Tsatsou, et al: Hybrid (content-based + domain ontology) | Advertisement (short text, video, ads, article) | Reference ontology, vocabulary (domain ontology), | Reasoning over the domain ontology, fuzziness and weighting, statistical analysis | Cold start problem (Relies on previous user ratings), static user profile |
| N. Alani, et al: Collaborative based | Academic activities and events | Reference ontology | Classification, concept relationship traversing | Unable to analyze implicit relations among community members |
| L. Zhuhadar, et al: Hybrid (content-based, Ruled-based) | Online Digital Corpus (HyperManyMedia) | Domain Ontology+ Learner's Ontology | Clustering+ Rule-based analysis among the user profile Hierarchy | Offline analysis of user interest |

## 6. CHALLENGES AND POTENTIAL IMPROVEMENTS

From the discussion above, we conclude that user contextual information is one of the key challenges for accurate personalization. A system which does not know about the user's context and user's goals often returns very general information in respond of user queries [34]. There is no commonly accepted definition of "context" in the personalization community [35]. But, we here refer to the most cited definition of context [35] as it is: "A context is any information that can be used to characterize the situation of an entity such as a person, place, or object. Thus, a good start of improvement performance of recommender system can be incorporating intensive contextual knowledge into the recommender system.

Considering the user situation in the context of adaptive systems, the following features can be identified: First, the interaction between a user and the system requires modelling the real-time system adaption based on the user goals, user prior knowledge, and user attitude with the prior results. For instance, user preference may change over the time depends on different situation. Second, the classification users into groups of similar interests need multiple and case-sensitive criteria than typical information such as demographic properties [38]. For example, users in the same groups based on age, education, and nationality have certainly different psychological characteristics that influence the object rating differently even in the same demographic-based classes. This situation is true in education systems that object rating is accomplished based on complex criteria such as cognitive properties [39] and mental perceptions of learners. Third, key properties of items, the features that specify objects in the given context, should be uncovered to the users for more accurate and rational rating. As an example, quality factors of learning objects in education domain such as books and lectures affects users decision to rate items. From the learner point of view, learning objects should be judged based on real discriminators features. Lastly, annotation and social tagging [40] attached on complex entities such as movies and e-books enables user to share object ratings can be engaged as a rich source of knowledge for efficient recommendation. Members of a social networks share their individual common sense on objects in a domain. However, trust and reputation issues [41] among communities in social networks should be considered.

Moreover, there are many other influential factors describing user context including user's short-term information needs, semantic knowledge about the domain, and user's long-term interests [21]. The effectiveness of recommender systems depends mainly on the completeness and accuracy of knowledge maintained in the user profiles. Many researches had shown that semantic approaches are effectively able to model user interests and increase recommendation accuracy [16][35][38]. Accordingly, we propose development of recommendation frameworks by incorporating new objects such as deep contextual features, user's cognitive properties, and case-sensitive mutli-criteria ratings into the recommendation process in order to increase the accuracy and performance of recommender systems.

## 7. CONCLUSION

Recommender systems assist users in finding more relevant information over a large volume of domain information by applying appropriate filtering techniques. User profile among other RS components plays an important role in recommender systems, and therefore has significant effects on the accuracy of recommendations because it provides the essential knowledge for computing user or item similarities, ranking, and making prediction about the user needs. The more effective user profiles have been employed, the more accurate recommendation and interesting items the user will receive.

This paper reveals that the integration of knowledge, specifically semantic information, into user profile not only improves the recommendation process but also tackles the inherent drawbacks of the respective approaches. The state-of-the art recommender systems approaches in terms of knowledge impression are discussed. It also represented how different type of knowledge is employed to overcome recommender drawbacks and to increase the recommendation accuracy. However, the type of knowledge, applicability of existing approach to the domain problem, and usefulness of knowledge are factors that affecting the task of incorporating knowledge into recommender system.

Furthermore, there is a trend in the recommendation community to exploit more semantic knowledge from the user context and user behaviour for constructing user profile which generally relies on extracting knowledge from the user context such as domain content, user interaction patterns, user cognitive properties, mutli-criteria, and social tagging. Therefore, more extensions in the recommendation area may put emphasis on incorporating new sort of knowledge such as sophisticated contextual knowledge, cognitive information, and background knowledge for increasing the performance of recommendation process. Thus, the alternative contribution of this paper is the proposal of engaging complementary knowledge into the user profile from user context by employing context-aware personalization approaches and context feature engineering methods that potentially improve the recommendation tasks.

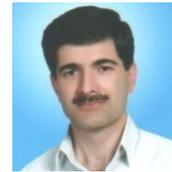

**Bahram Amini** received his B.Sc. and M.Sc. degrees in Software Engineering from Tehran University and Azad University respectively in Iran. He has 14 years of education and research experience in academics and industry at internationally repute organizations in Iran. He is now a Ph.D. student at the Faculty of Computer Science and Information Systems at UTM, Malaysia. He is a member of the Society of Digital Information and Wireless Communications (SDIWC) and Applied Ontology and Conceptual Modeling Special Interest Group (AOCO-SIG). He has authored a number of research papers and articles for various national and international Journals/conferences. His current research interests include web intelligent, adaptive systems, text mining, and web services.

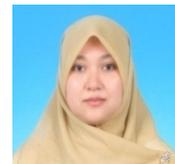

**Dr. Roliana Ibrahim** is a senior lecturer at the Faculty of Computer Science and Information Systems, Universiti Teknologi Malaysia. She received her Doctoral Degree in Information Science from Loughborough University, UK. She is also a member or Software Engineering Research Group (SERG) and currently involves in project relating to the development of ontology-based data repository and data mining algorithm to mine heterogeneous data for cancer research. Her current research interests are the adoption of system thinking methodologies and ontology for complex systems integration and development, data warehousing and data mining. She is also an Oracle Certified Associate and hold certificates as Microsoft Certified Professional Developer (MCPD) in ASP.NET Developer and Microsoft Certified Technology Specialist (MCTS) in SQL Server 2008, Database Development.

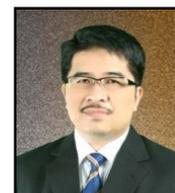

**Dr. Mohd Shahizan Othman** received his BSc in Computer Science with a major in Information Systems from Universiti Teknologi Malaysia (UTM), Malaysia, in 1998. Then he earned Msc in Information Technology from the Universiti Kebangsaan Malaysia (UKM), Malaysia. Soon after, he graduated for his PhD in Web Information Extraction, Information Retrieval and Machine Learning from UKM. He is currently a senior lecturer at the Faculty of Computer Science and Information Systems, UTM, since 2001. His research interests are in information extraction and information retrieval on the web, web data mining, content management and machine learning.